\documentclass[aps,prl,letterpaper,10pt, twocolumn,showpacs,superscriptaddress]{revtex4-2}

\usepackage{hyperref}
\usepackage{amssymb}
\usepackage{amsmath}
\usepackage{graphicx}
\usepackage{subfigure}
\usepackage{amsfonts}
\usepackage{xcolor}
\usepackage{epsfig}
\usepackage{color}
\usepackage{bbm}

\newcommand{\m }{\mathcal}

\newcommand{\bra}[1]{\langle#1|}
\newcommand{\ket}[1]{|#1\rangle}

\newcommand{\mean}[1]{\langle #1 \rangle}
\newcommand{\proj}[1]{\ket{#1}\!\bra{#1}}

\begin{document}
	
	
\title{Bosonic Indistinguishability-Dependent Contextuality}


\author{Ali Asadian}
\email{ali.asadian668@gmail.com}
\affiliation{Department of Physics, Institute for Advanced Studies in Basic Sciences (IASBS), Gava Zang, Zanjan 45137-66731, Iran}

\author{Ad\'an~Cabello}
\email{adan@us.es}
\affiliation{Departamento de F\'{\i}sica Aplicada II,
	Universidad de Sevilla,
	E-41012 Sevilla,
	Spain}
\affiliation{Instituto Carlos~I de F\'{\i}sica Te\'orica y Computacional, Universidad de
	Sevilla, E-41012 Sevilla, Spain}
\date{\today}


\begin{abstract}
We uncover a form of quantum contextuality that connects maximal contextuality to boson indistinguihability in a similar way maximal nonlocality with respect to the Clauser-Horne-Shimony-Holt Bell inequality is connected to maximal entanglement. Unlike previous forms of photonic contextuality, this form cannot be simulated with classical light, as it relies on indistinguishability and higher-order interference. Ideal measurements on the bosonic system can be performed by means of dispersive coupling with an ancillary qubit. This allows us delaying at will the ending of each measurement and targeting high-dimensional contextual correlations, which are features which cannot be achieved with existing platforms.
\end{abstract}


\maketitle


{\em Introduction.---}Kochen-Specker contextuality \cite{Specker60,KS67,Bell66} is a fundamental property of quantum mechanics. Intuitively, a measurement $M$ that yields the same result when repeated and does not disturb any compatible observable must be revealing a predetermined result that is independent of the ``context,'' defined as the set of compatible observables measured along with~$M$. Measurements with these properties are called ideal, sharp, or projective \cite{CY16,Cabello19,PZHCKH20}. In contrast, in quantum mechanics, sequences of ideal measurements \cite{KZG09,ZUZ13,LMZNCAH18}
can produce correlations which cannot be explained assuming that results are noncontextual \cite{KCBS08,Cabello08,BBCP09,YO12,KBLGC12}. This phenomenon is manifested in the violation of noncontextuality (NC) inequalities \cite{KCBS08,Cabello08,BBCP09,YO12,KBLGC12}, which must be satisfied by any noncontextual model. Violations of these inequalities can be observed with local measurements on spatially separated subsystems, as in Bell tests \cite{Bell64}, and in experiments with sequential measurements on single systems \cite{KZG09,ARBC09,ZUZ13,AACB13,DHANBSC13,MANCB14,LMZNCAH18,MZLCAH18}. Kochen-Specker contextuality has multiple applications \cite{BCGKL21} and play a fundamental role in quantum computation \cite{HWVE14,DGBR15,RBDOB17} and quantum foundations \cite{Cabello13,Cabello19}.

The present work is motivated by two observations. One is that, while maximal nonlocality with respect to the Clauser-Horne-Shimony-Holt (CHSH) Bell inequality \cite{CHSH69} requires maximal bipartite entanglement \cite{MY04}, so far, no link has been established between maximal single-particle Kochen-Specker contextuality and a fundamental resource.
 
The second observation is that the contextual correlations in experiments with sequential measurements on single photons in interferometric setups (e.g., \cite{ARBC09,MANCB14}) can be simulated with classical light \cite{FBV16,ZXX19} since measurements are first-order coherence measurements and thus probability distributions can be associated to single-mode intensities \cite{FBV16}. This classical simulability contrasts with the fact that both quantum computing with linear optics \cite{KLM01} and boson sampling \cite{AA11,ZWD20} crucially rely on bosonic indistinguishability and higher-order interference. 

Therefore, a question naturally arises: Is there a form of Kochen-Specker contextuality linked to boson indistinguishability and higher-order coherence measurements?
This is an open problem despite contextuality for bosonic systems has been discussed in \cite{CT13,Cabello12,KSTK14}.

In this work, we describe a method to produce contextual correlations between sequences of compatible ideal measurements on systems of a fixed number of identical bosons in such a way that indistinguishability is necessary for maximal contextuality with respect to several fundamental NC inequalities. Hence, we will refer to this effect as bosonic indistinguishability-dependent contextuality (BIC).



{\em BIC with two identical bosons.---}Contextuality for ideal measurements requires quantum systems of dimension three (qutrits) or higher \cite{Specker60,KS67,Bell66}. For this reason, we begin by considering the following case: two indistinguishable photons propagating through two spatially distinct modes (upper and lower). This system can encode a qutrit, since an orthogonal basis is $\{|2,0\rangle, |0,2\rangle,|1,1\rangle\}$, where $|n_a,n_b\rangle$ denotes the state with $0 \le n_a \le 2$ photons in the upper mode and $n_b=2-n_a$ photons in the lower mode. It is worth mentioning that $\ket{2,0}$ and $\ket{0,2}$ indicates that two identical bosons can occupy the same mode, the property which is sometimes refers to bosonic bunching. 

For witnessing contextuality, we will consider the most fundamental NC inequality for qutrits, the Klyachko-Can-Binicio\u{g}lu-Shumovsky (KCBS) inequality \cite{KCBS08}, which can be written as follows:
\begin{equation}
	\label{K}
	\kappa = - \frac{1}{3} \sum_{j=1}^5 \langle A_j A_{j+1} \rangle \le 1,
\end{equation}
where $A_j$ are observables with possible results $-1$ and $+1$, $ \langle A_j A_{j+1} \rangle$ is the mean value of the product of the results of $A_j$ and $A_{j+1}$, and the sum is taken modulo $5$. An experimental test of the violation of the KCBS inequality with sequential ideal measurements requires preparing a particular initial qutrit state $\ket{v}$ and then performing two sequential compatible ideal measurements of the type $A_j=2 \proj{v_j}-\openone$.

To obtain the initial qutrit state needed, we allow the two modes interacting in a beam splitter (BS). This produces the following input-output transformation \cite{SM}:
\begin{subequations}
	\begin{align}
		\label{eq:BeamTrans1}
		U_{\rm BS}(\theta, \phi)a U_{\rm BS}^\dag(\theta, \phi)&= \cos (\theta/2) a- e^{i\phi}\sin (\theta/2) b, \\
		\label{eq:BeamTrans2}
		U_{\rm BS}(\theta, \phi)b U_{\rm BS}^\dag(\theta, \phi)&= e^{-i\phi}\sin (\theta/2) a+\cos(\theta/2) b,
	\end{align}
\end{subequations}
where $a$ ($a^\dag$) and $b$ ($b^\dag$) are the annihilation (creation) operators for the upper and lower mode, respectively, and $\theta$ and $\varphi$ are the angles accounting for the transmitivity and phase shift introduced by the BS, respectively. 

If one begins with state $|1,1\rangle=a^\dag b^\dag\ket{\rm vac}$, where $\ket{\rm vac}$ denotes the vacuum state, then the BS produces \cite{SM}
\begin{equation}
		\label{11}
		U_{\rm BS}(\theta, \phi) \ket{1,1}=
		\tfrac{\sin\theta e^{i \phi}}{\sqrt{2}}\ket{2,0}
		- \tfrac{\sin\theta e^{-i\phi}}{\sqrt{2}}\ket{0,2}
		+\cos\theta \ket{1,1}.
\end{equation}
By suitably choosing $\theta$ and $\phi$, one can produce the bosonic analog of any qutrit state $\ket{v}$ with real components $(v_x,v_y,v_z)$ in a Cartesian basis. This follows from the fact that $\ket{v}$ can be written as $(v_+,v_-,v_0)$ in a spherical basis, where $v_{\pm}=(\mp v_{x}+i v_{y})/\sqrt{2}$ and $v_0=v_z$. Using that $v_x=\sin\theta \cos\phi$, $v_y=\sin\theta \sin\phi$, and $v_z=\cos\theta$, $(v_+,v_-,v_0)$ correspond to the components in the basis $\{|2,0\rangle, |0,2\rangle,|1,1\rangle\}$, respectively, as shown in Eq.~\eqref{11} \cite{SM}. 

\begin{figure}[!]
	\includegraphics[width=\linewidth]{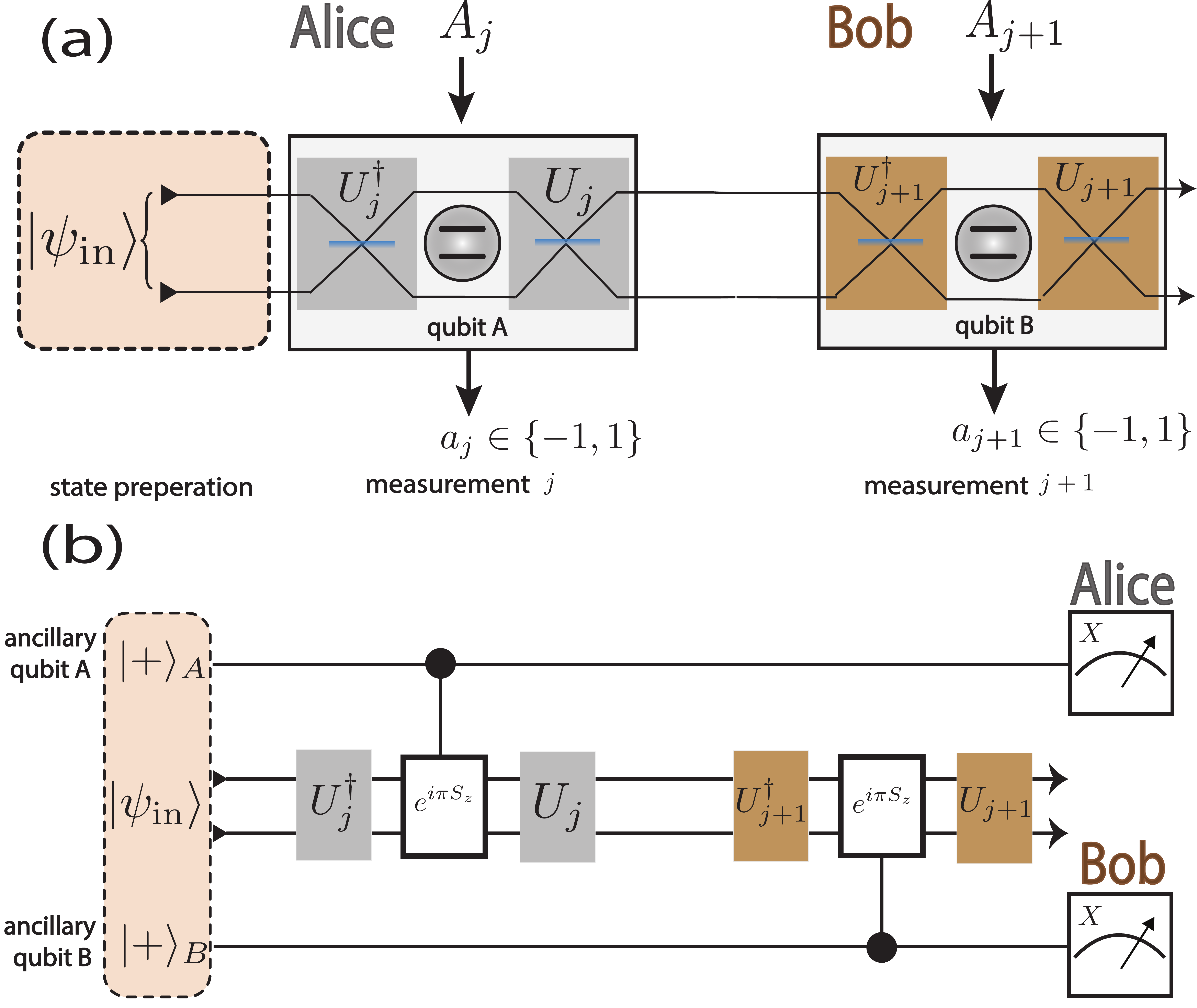}
	\caption{(a) Schematic setup of two sequential ideal measurements on a initial state of a two-mode two-photon system. (b)~Circuit representation of the operations used for the two sequential ideal measurements. Each of the measurements is finished only when the corresponding party reads out its qubit.}
	\label{Fig1}
\end{figure}

Therefore, for preparing $|\psi_{\text{in}}\rangle$, defined as the bosonic analog of a qutrit state $\ket{v}$, one can start with $|1,1\rangle$, which is easy to prepare by pumping a nonlinear crystal with an intense laser to generate pairs of identical photons \cite{BW70,Kwiat95}, and then apply a suitably chosen BS. This is what is meant to happen in the block ``state preparation'' in Fig.~\ref{Fig1} (a). 

The choice of observables and the procedure for performing ideal measurements of them are crucial for connecting contextuality to bosonic indistinguishability. 
In general, an ideal measurement is implemented by applying first a suitable unitary transformation, then performing a nondemolition basic measurement, and then applying the adjoint unitary transformation. For the unitary transformations, we use BSs: one before the measurement, implementing 
$U^\dag_{\rm BS}(\theta_j, \phi_j)$, and one after the measurement, implementing $U_{\rm BS}(\theta_j, \phi_j)$; see Fig.~\ref{Fig1}, where $U_j\equiv U_{\rm BS}(\theta_j, \phi_j)$.

For choosing a basic measurement, notice that, for state \eqref{11}, the probability for detecting one photon in each of the output ports of the BS is $p_+=\cos^2\theta$ and the probability of observing photon bunching is $p_-=1-p_+$. In the case of a balanced BS (i.e., when $\theta=\pi/2$), $p_-=1$, as was first demonstrated by Hong, Ou, and Mandel (HOM) \cite{HOM87}. Therefore, for our purpose, a natural choice for the basic measurement is the one in which outcome~$+1$ corresponds to the two photons being detected in coincidence at the two modes, and outcome~$-1$ to the bunching of the two photons. Notice that, in our case, the two photons are noninteracting thus the bunching is originated from their indistinguishablity. This basic measurement is represented by the projector $|1,1\rangle\langle1,1|$, associated to outcome~$+1$, and the projector $\mathbbm{1}-|1,1\rangle\langle1,1|$, associated to outcome $-1$. Therefore, the measurement corresponding to what is represented by $\proj{v_j}$ in the qutrit case, is represented in our setup by $U_{\rm BS}(\theta_j, \phi_j) \proj{1,1} U_{\rm BS}^\dag(\theta_j, \phi_j)$. 
The inner product between a pair of vectors is $\bra{v_i}v_j\rangle=\cos\theta_i\cos\theta_j+\sin\theta_i\sin\theta_j\cos(\phi_i-\phi_j)$ from which the suitable BS angles associated to the $A_j$ and $A_{j+1}$ are determined. The maximum violation is achieved for $\bra{v_j}\psi_{\rm in}\rangle=\bra{v_{j+1}}\psi_{\rm in}\rangle=\cos\gamma$, where $\gamma=\cos^{-1} (1/5^{1/4}) $.
In this case, one obtains $\bra{\psi_{\rm in}}A_j A_{j+1}\ket{\psi_{\rm in}}=-4\cos^2\gamma +1$, which
gives
\begin{equation}
	\label{max}
	\kappa = \frac{4 \sqrt{5}-5}{3} \approx 1.315,
\end{equation}
which is the highest possible violation of inequality \eqref{K} allowed by quantum theory with ideal measurements \cite{CSW14}. This value can be achieved by choosing $\theta=0$ in the BS of the state preparation (thus $\ket{\psi_{\rm in}}=\ket{1,1}$) and choosing the angles of the BSs
for the observables $A_j$ as follows: 
$\theta_j=\cos^{-1} (1/5^{1/4})$ and $\phi_j=2\pi j/5$.


{\em Scheme for ideal measurements.---}For Kochen-Specker contextual correlations between sequential measurements, we need the measurements to be ideal. That is, repeatable and nondisturbing compatible observables that may be measured afterwards. The detection of the photons would kill the possibility of performing further measurements. Therefore, we need an interaction, between the two-mode two-photon system and an ancillary system, which encodes the result of the measurement in the ancilla and leaves the bosonic system in the post-measurement state given by L\"uders' rule \cite{PZHCKH20}. 

The dispersive regime \cite{Gao18,Gao19,Magnard20,CQED} provides an efficient method to do it without destroying the photon number state. 
The required Hamiltonian is
\begin{equation}
	H_{\rm disp}=\lambda(t) S_z\otimes\sigma_z,
\end{equation}
with $S_z=(a^\dag a-b^\dag b)/2$ and $\sigma_z=\proj{\uparrow}-\proj\downarrow$, where $\ket{\uparrow}$ and $\ket{\downarrow}$ are the basic states of the ancilary qubit. 

The interaction is engineered in such a way that $\int_0^\tau dt \lambda(t)=\pi$. Therefore, $U_{\rm disp}(\pi) =e^{i\pi S_z\otimes \sigma_z}$. This has the following effect:
\begin{subequations}
	\begin{align}
		U_{\rm disp}(\pi)\ket{2,0}\ket{+}&=\ket{2,0}e^{i\pi\sigma_z}\ket{+}= \ket{2,0}\ket{-}, \\
		U_{\rm disp}(\pi)\ket{0,2}\ket{+}&=\ket{0,2}e^{-i\pi\sigma_z}\ket{+}=\ket{0,2}\ket{-}, \\
		U_{\rm disp}(\pi)\ket{1,1}\ket{+}&=\ket{1,1}\ket{+},	
	\end{align}
\end{subequations}
where $\ket{\pm}=(\ket{\downarrow}\pm \ket{\uparrow})/\sqrt{2}$ are states of the ancillary qubit.
By using this dispersive coupling, we can encode the outcomes of our basic measurement in the state of the ancillary qubit. 
The measurement outcome can be later read out by performing projective measurement on the ancillary qubit. 
These types of couplings and measurements are currently efficiently implemented in various cavity and circuit quantum electrodynamics setups \cite{Gao18,Gao19,Magnard20,CQED}.

Similarly, the measurement of any of the required observables involves $U_{A_j}=U_{\rm BS}(\theta_j, \phi_j) U_{\rm disp}(\pi) U^\dag_{\rm BS}(\theta_j, \phi_j)$ acting on the initial state,
\begin{align}
	U_{A_j}\ket{\psi_{\rm in}}\ket{+} &=\cos\gamma\ket{v_j}\ket{+}+\sin\gamma\ket{v_j^\bot}\ket{-} \nonumber \\
	&= K_+(j)\ket{\psi_{\rm in}}\ket{+}+K_-(j)\ket{\psi_{\rm in}}\ket{-},
\end{align}
where $K_\pm(j)=(\mathbbm{1}\pm \m P_j)/2$ are Kraus operators acting on the bosonic modes subspace. $\m P_j=U_{\rm BS}(\theta_j, \phi_j) e^{i\pi S_z} U^\dag_{\rm BS}(\theta_j, \phi_j)$. The outcome probabilities corresponding to the qubit measurements on the $\{\ket{+},\ket{-}\}$ basis are $p_\pm=\bra{\pm}\rho_A \ket{\pm}=\bra{\psi_{\rm in}}E_\pm(j)\ket{\psi_{\rm in}}$, where $E_\pm(j)=K^\dag_\pm(j)K_\pm(j)=(\mathbbm{1}\pm \m A_j)/2$ are the associated positive operator-valued measure (POVM) with $\m A_j=(\m P_j+\m P_j^\dag)/2$. Therefore,
\begin{equation}
	\mean{\sigma_x}=p_+-p_-=\bra{\psi_{\rm in}}\m A_j \ket{\psi_{\rm in}} ,
\end{equation}
where, $p_+=\cos^2\gamma$ and $p_-=1-p_+$.

If the total number of photons in the two modes is even, corresponding to an odd-dimensional Hilbert space, then $\m P_j$ is a Hermitian operator with eigenvalues $\pm 1$.
For the two-photon case the observable reduces to $\m A_j=\m P_j$, defined in a three-dimensional Fock space.

The crucial requirement is that the interaction between the photons and the qubit should preserve the indistinguishability of the input photons. 
The dispersive interaction induces $\sigma_z$ eigenstate-dependent frequency shift to the both modes, i.e., $\omega_{a(b)}=\omega\pm\lambda$. However, the
qubit initialized in $\ket{+}$ does not induce a frequency shift.
The second measurement is similar. 

This measurement technique enables sequential ideal measurements on the two-mode two-photon system.
The joint probabilities are
\begin{align}
	p_{a_j a_{j+1}}=&\bra{\psi_{\rm in}}K^\dag_{a_j}(j) E_{a_{j+1}}(j+1)K_{a_j}(j)\ket{\psi_{\rm in}},
\end{align}
where $a_j, a_{j+1}\in\{-,+\}$. Once $p_{a_j a_{j+1}}$ are plugged into the mean values we get,
\begin{align}
	{\rm tr}\Big(\rho_{AB}(j)\sigma^A_x\otimes\sigma^B_x\Big)&\equiv\bra{\psi_{\rm in}}\m A_j \m A_{j+1}\ket{\psi_{\rm in}}.
\end{align}
The initial state of the two ancilla qubits $\ket{+}_A\ket{+}_B$ after the action of $U_{A_{j+1}}U_{A_j}$ turns into $\rho_{AB}(j)= p_{--}\proj{--}+(1-p_{--})\proj{\psi^+}$, where $\ket{\psi^+}=(\ket{+-}+\ket{-+})/\sqrt{2}$. Note that $p_{++}=0$. This is due to the bosonic bunching or HOM-like effect yielding $p_{+|+}=|\bra{1,1}U^\dag_j U_{j+1}\ket{1,1}|^2=0$, as a direct cosequence of bosonic indistinguishability. 
The propagating bosonic system acts as a quantum bus \cite{bus}, entangling the two ancillary qubits. Therefore, the correlations demonstrated by the two qubits cannot be generated by coupling to classical fields.


{\em Higher-dimensional BIC correlations.---}The method described before can be used to produce the BIC analogs of the following quantum contextual correlations with qutrits: any quantum violation of any of the generalizations of the KCBS inequality for odd $n$-cycle scenarios (which are the only tight NC inequalities for these scenarios) \cite{CDLP13,AQB13,Cabello:2016JPA,MZLCAH18}, the Yu-Oh inequality and its optimal versions (which are the simplest tools for witnessing state-indepent contextuality) \cite{YO12,KBLGC12,ZUZ13,Cabello:2016JPA}, the NC inequalities for Kochen-Specker sets in dimension three \cite{BBCP09,Peres:2006Book}, and the Hardy-like proof of contextuality \cite{CBTB13,MANCB14}.

However, there are quantum contextual correlations of interest that can only be achieved using high-dimensional quantum systems (see, e.g., \cite{CDLP13,Cabello13b,LBPC14,ATC15}) and have potential applications (e.g., in dimension witnessing \cite{Guhne:2013PRA,Ray:2021NJP}, self-testing \cite{Kishor,Bharti:2019XXX}, and machine learning \cite{Gao:2021XXX}). The problem is that, so far, the largest quantum system on which sequential ideal measurements have been carried out has dimension four \cite{KZG09} and current platforms do not offer a real chance to produce correlations between sequential ideal measurements on high-dimensional systems.

This points out one of the potential applications of BIC: producing high-dimensional contextual correlations with ideal measurements. The method presented can be extended to the case of $m$-mode $n$-photon systems. Current technology allows for preparing bosonic systems with up to $m=100$ and $n=76$, yielding a state space dimension of about $10^{30}$ \cite{ZWD20}. On the other hand, dichotomic ideal measurements can be carried out applying the same interaction with a qubit described before but on a $m>2$-mode $n>2$-photon systems as in \cite{HD0,HD1,HD2,HD3}. It is important to recall that any matrix of quantum contextual correlations can be produced in an experiment consisting of sequences of {\em two} dichotomic measurements \cite{Cabello16}. Moreover, these experiments only require one ideal measurement (the first one), since the second can be a destructive measurement. 

While realizing the unitaries needed for achieving specific high-dimensional correlations may require a specific analysis,
the method presented above can be extended in a straightforward way to two-mode $n$-photon systems, since, due to the Jordan-Schwinger map \cite{Jordan35,Schwinger52}, they allow for a bosonic realization of a qudit of dimension $d=n+1$. A generalization to $m$-mode $n$-photon systems allowing $d=\frac{(n+m-1)!}{n!(m-1)!}$ is also possible.



\begin{figure}[!]
	\includegraphics[width=\linewidth]{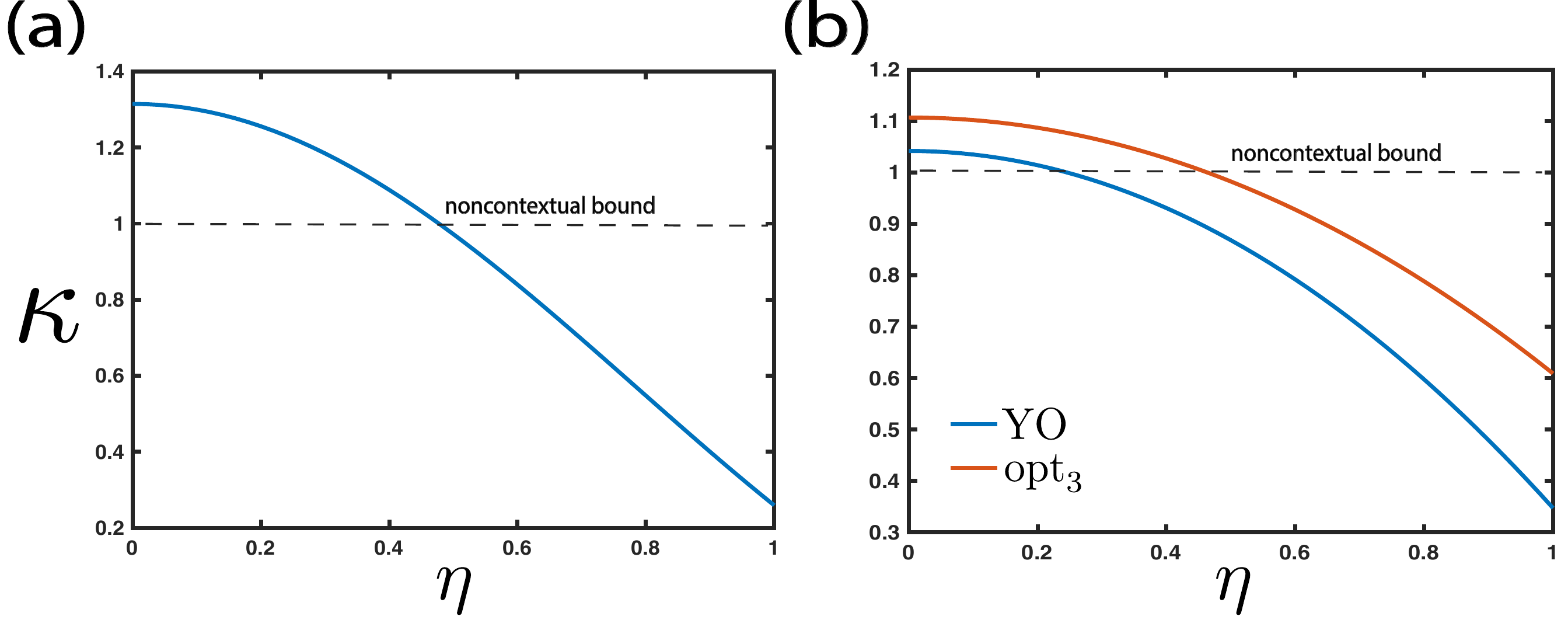}
	\caption{ Value of the contextuality witness (whith noncontextual $1$) as a function of the degree of bosonic indistinguishability $\eta$, defined in the text. (a)~For the witness defined in \eqref{K}, with initial states $U_{\rm BS}(\theta, \phi) \ket{1,1_\eta}$, defined in the text, and the observables $A_j$ used to obtain (\ref{max}). (b)~For the Yu-Oh and the optimal state-independent inequality opt$_3$ witnesses \cite{SM}, for {\em any} initial state $U_{\rm BS}(\theta, \phi) \ket{1,1_\eta}$ (i.e., no matter the values of $\theta$ and $\phi$).}
	\label{Fig2}
\end{figure}


{\em State-independent BIC.---}One of the interesting possibilities of quantum contextuality is that it can be state independent, which means that, for any quantum system of dimension three or larger, there are sets of measurements and contextuality witnesses that have the same value (beyond the corresponding noncontextual bound) for any quantum state \cite{Cabello08,BBCP09,YO12,KBLGC12}. The same happens for BIC.

To show this effect, one can consider the bosonic equivalents of the witness of Yu and Oh, YO \cite{YO12}, or its optimal version, opt$_3$ \cite{KBLGC12}, and apply the method described before. As it can be easily checked, while the noncontextual bound for both witnesses is $1$, any state in the basis $\{|2,0\rangle, |0,2\rangle, |1,1\rangle\}$ gives the value $25/24 \approx 1.042$ for YO and the value $83/75 \approx 1.107$ for opt$_3$ \cite{SM}.


{\em Maximum contextuality requires perfect indistinguishability.---}So far, we have assumed that the two photons in the modes are indistinguishabile. Here, we study what happens when they are not perfectly indistinguishable. This may occur due to, e.g., that they have a different polarization or that there is a time delay between them. 

To model the effect of distinguishability, one can replace the initial state $\ket{1, 1}$, with which the state preparation was fed, with the state $\ket{1, 1_\eta}=a^\dag b_\eta^\dag\ket{\rm vac}$ with $b^\dag_\eta=\sqrt{1-\eta^2} b^\dag +\eta b_\bot^\dag$, where $a^\dag$ and $b^\dag$ are creation operators of indistinguishable photons in the upper and lower modes, respectively, while $b^\dag_\bot$ is a creation operator of photons in the lower mode which are perfectly distinguishable form the former (e.g., $a^\dag$ and $b^\dag$ create horizontally polarized photons and $b^\dag_\bot$ creates vertically polarized ones). $\eta\in[0,1]$ quantifies the degree of distinguishablity between the two photons, with $\eta=0$ representing perfect indistinguishability and $\eta=1$ perfect distinguishability. 

The BS mixes the upper and lower modes regardless of how distinguishable the photons are. Therefore, we are dealing with four distinct modes then, i.e., thoses corresponding to $a$, $a_\bot$, $b$, and $b_\bot$.

Assuming that the dispersive coupling merely depends on the number of photons in each transmission line and not on the degree of freedom with respect to which the photons are distinguishable,
\begin{equation} 
	H_{\rm disp}=\lambda(t)\frac{(N_a- N_b)}{2}\otimes\sigma_z=\lambda(t)(S_z+S^\bot_z)\otimes \sigma_z,
\end{equation}
where
$N_a=a^\dag a+ a^\dag_\bot a_\bot$ denoting the number of photons in the upper transmission line and $N_b=b^\dag b+ b^\dag_\bot b_\bot$ denoting the number of photons in the lower transmission line. Fig.~\ref{Fig2}(a) shows how the value of $\kappa$ depends on $\eta$: The maximum value is only obtained when the photons are perfectly indistinguishable. Otherwise, contextuality descreases as distinguishability increases. 
This implies that contextuality can be used to certify boson indistinguishability in a way that cannot be simulated with classical light. This contrasts with the fact that the HOM effect can be simulated with classical light \cite{HOMS}. Moreover, since the maximum quantum violation of the KCBS inequality allows for self-testing \cite{Kishor} (i.e., certification using only the observed correlations), Fig.~\ref{Fig2}(a) shows that BIC can be used to self-test boson indistinguishability. Therefore, BIC provides a quantitative test of quantum indistinguishability much more robust and detailed than the one provided by a HOM setup. 

Interestingly, Fig.~\ref{Fig2}(b) shows exactly the same behavior, but now the degree of contextuality does not depend on $\theta$ and $\phi$ of the initial state, but only on the degree of distinguishability $\eta$ \cite{SM}.


{\em Delayed measurements.---}The method for performing sequential measurements described above opens an interesting possibility: deciding at will the order in which the sequential measurements are finished. 
This possibility comes from the fact that, in the case of sequences of two measurements, the quantum state after the interaction with the second ancillary qubit is a coherent superposition of the four quantum states corresponding to the four combinations of results for the two measurements \cite{SM}. 
This superposition can be ``collapsed'' in three different ways: (i)~by first reading out the second qubit and only then reading out the first qubit, (ii)~by first reading out the first and then the second, or (iii)~by spacelike separating the readouts. This offers a possibility beyond what can be done in standard sequential measurement experiments (e.g., \cite{LMZNCAH18}), where the readout of the result of the second measurement cannot be spacelike separated from the readout of the result of the result of the first measurement. This possibility can stimulate a new generation of sequential measurement experiments and tests of collapse models \cite{Bassi13}.


{\em Conclusions.---}Photons are essential for quantum communication and contextuality is crucial for quantum speedup and secure communication. However, current forms of photonic contextuality can be simulated with classical light, which brings the question of whether there are other forms which cannot. Here, we have introduced a new form of photonic contextuality, dubed bosonic indistinguishability-dependent contextuality (BIC), that relies on indistinguishability and higher-order interference, connects maximum contextuality with perfect indistinguishability, allows producing high-dimensional contextual correlations between ideal measurements, and experiments with sequential meaurements in which the ending of each measurement can be delayed at will, features all of them which cannot be achieved in existing platforms. BIC may unblock the field of high-dimensional quantum contextuality and stimulate new tests of collapse models. In addition, BIC could push the search for connections between contextuality, photonic universal quantum computation, and boson sampling.


\begin{acknowledgments}
AA thanks J. Majer for useful discussion on the possible cQED implementation.
This work was supported by Project Qdisc (Project No.\ US-15097), with FEDER funds, MINECO Project No.\ FIS2017-89609-P, with FEDER funds, and QuantERA grant SECRET, by MINECO (Project No.\ PCI2019-111885-2).
\end{acknowledgments}


\appendix

\section*{Supplemental material}


\subsection{Action of a beam splitter on a two-mode two-photon system and generation of the bosonic equivalent of any qutrit state with real components}


The Hamiltonian of a beam splitter (BS) coupling the two input modes (upper and lower) is
\begin{equation}
	H_{\rm BS}=\frac{i \theta}{2}(e^{-i\phi} a^\dag b- e^{i\phi} b^\dag a),
\end{equation}
where $a^\dag$ and $a$ are the creation and annihilation operators for the upper mode, $b^\dag$ and $b$ are the creation and annihilation operators for the lower mode, and $\theta$ and $\varphi$ are the angles accounting for the transmitivity and phase shift introduced by the BS, respectively.
Therefore, the BS transformation is
\begin{equation}
	U_{\rm BS}(\phi,\theta)=\text{exp}\left[\frac{\theta}{2}( e^{-i\phi} a^\dag b- e^{i\phi} b^\dag a)\right].
\end{equation}
In the qutrit subspace spanned by the basis
$\{\ket{2, 0}, \ket{0, 2}, \ket{1,1}\}$, where $|n_a,n_b\rangle$ is the state in which there are $0 \le n_a \le 2$ photons in the upper mode and $n_b=2-n_a$ photons in the lower mode, the action of $U_{\rm BS}(\phi,\theta)$ is
\begin{equation}
	U_{\rm BS}(\phi,\theta)= \left( \begin{array}{ccc}
		\cos^2\dfrac{\theta}{2} & \dfrac{1}{\sqrt{2}}e^{i\phi}\sin\theta & e^{i2\phi}\sin^2 \dfrac{\theta}{2} \\
		-\dfrac{1}{\sqrt{2}}e^{-i\phi}\sin\theta& \cos\theta & \dfrac{1}{\sqrt{2}}e^{i\phi}\sin\theta \\
		e^{-i2\phi}\sin^2 \dfrac{\theta}{2} & -\dfrac{1}{\sqrt{2}}e^{-i\phi}\sin\theta & \cos^2\dfrac{\theta}{2} \end{array} \right),
\end{equation}
which applied to $\ket{1,1}$ produces the state given by Eq.~\eqref{11}.

By suitably chosen $\theta$ and $\phi$ of the BS, one can produce the bosonic analog of any $\ket{v}$ with real components. This can be seen as follows. Consider
\begin{equation}
	\ket{v}=v_x \ket{e_x}+v_y \ket{e_y}+v_z \ket{e_z},
\end{equation}
where $(v_x,v_y,v_z) \in \mathbb{R}^3$ and $\{\ket{e_x}, \ket{e_y}, \ket{e_z}\}$ is a Cartesisan basis.
Consider the spherical basis, defined as
\begin{subequations}
\begin{align}
	&\ket{e_\pm} = \mp\tfrac{1}{\sqrt{2}}(\ket{e_x}\pm i \ket{e_y}),\\	
	&\ket{e_0} = \ket{e_z}, 
\end{align}
\end{subequations}
where $i$ denotes the imaginary unit. Then,
\begin{equation}
	\ket{v}=v_- \ket{e_-}+v_+ \ket{e_+}+v_0 \ket{e_0}.
\end{equation}
The components in the spherical basis are related to the components in the Cartesian basis by
\begin{subequations}
\begin{align}
	&v_\pm = \tfrac{1}{\sqrt{2}}(\mp v_x+ i v_y),\\
	&v_0 = v_z.
\end{align}
\end{subequations}
Now notice that $\ket{v}$ can be written as
\begin{equation}
	\ket{v}=\sin\theta\cos\phi \ket{e_x} + \sin\theta\sin\phi \ket{e_y} + \cos\theta \ket{e_z},
\end{equation} 
with $0 \le \theta < \pi$ and $0 \le \phi < \pi$.
Therefore, the components of $\ket{v}$ in the spherical basis are
\begin{subequations}
\begin{align}
	&v_+=\tfrac{1}{\sqrt{2}}(-\sin\theta\cos\phi + i \sin\theta\sin\phi)=-\tfrac{\sin\theta}{\sqrt{2}}e^{-i\phi}, \\
	&v_-=\tfrac{1}{\sqrt{2}}(\sin\theta\cos\phi + i \sin\theta\sin\phi)=\tfrac{\sin\theta}{\sqrt{2}}e^{i\phi}, \\
	&v_0 = \cos\theta,
\end{align}
\end{subequations}
which, as shown in Eq.~\eqref{11}, are in one-to-one correspondence with the components of $U_{\rm BS}(\theta, \phi)\ket{1,1}$ in the basis $\{\ket{2,0},\ket{0,2},\ket{1,1}\}$, respectively. The correspondence is: $\ket{0,2}\Leftrightarrow \ket{e_+}$, $\ket{2,0}\Leftrightarrow \ket{e_-}$, and 
$\ket{1,1}\Leftrightarrow \ket{e_0}$.


\subsection{Quantum state after the interaction with two successive ancillary qubits}

	
The state of the bosonic system, the ancillary qubit for the first measurement, and the ancillary qubit for the second measurement, before the readout of the two qubits, is
\begin{align}
\label{sout}
 \ket{\psi_{\rm out}}=&U_{A_{j+1}}U_{A_j}\ket{\psi_{\rm in}}\ket{+}_A\ket{+}_B \nonumber \\
 =&\sqrt{p_{++}} \ket{v_{j+1}}\ket{+}_A\ket{+}_B +\sqrt{p_{+-}} \ket{v^\bot_{j+1}}\ket{+}_A\ket{-}_B \nonumber \\
 &+\sqrt{p_{-+}} \ket{v_{j+1}}\ket{-}_A\ket{+}_B +\sqrt{p_{--}} \ket{v^\bot_{j+1}}\ket{-}_A\ket{-}_B,
\end{align}
where,
\begin{subequations}
\begin{align}
	p_{++}&=|\bra{v_j}v_{j+1}\rangle|^2 \cos^2\gamma , \\ p_{+-}&=(1-|\bra{v_j}v_{j+1}\rangle|^2) \cos^2\gamma,\\
	p_{-+}&=|\bra{v_j}v^\bot_{j+1}\rangle|^2 \sin^2\gamma ,\\ p_{--}&=(1-|\bra{v_j}v^\bot_{j+1}\rangle|^2) \sin^2 \gamma,
\end{align}
\end{subequations}
with
\begin{equation}
	\ket{v^\bot_j}=\sin\theta\ket{1,1} -\cos\theta \frac{e^{i\phi_j}\ket{2,0} -e^{-i\phi_j}\ket{0,2}}{\sqrt{2}}.
\end{equation}
State \eqref{sout} is a coherent superposition of the four distinct combinations of the measurement results for $A_i$ and $A_{j+1}$. Therefore, the order in which the two sequential measurements are finished is determined by the order in which the ancillary qubits are readout.


\subsection{Witnesses for state-independent contextuality}


The Yu-Oh inequality \cite{YO12} can be written as
\begin{equation}
	\text{YO}=-\frac{1}{8}\left( \sum_{j\in V} \langle A_j \rangle + \frac{1}{2} \sum_{(j,k) \in E} \langle A_j A_k \rangle \right) \le 1,
\end{equation}
where $A_j$ are $13$~observables with possible results $-1$ and $+1$, $V=V_1 \cup V_2$, with $V_1=\{1,2,\ldots,9\}$ and $V_2=\{A,B,C,D\}$, and $E=E_1 \cup E_2$, with $E_1=\{(1,4),(1,7),(2,5),(2,8),(3,6),(3,9),(4,7),(5,8),$ $(6,9)\}$ and $E_2=\{(1,2),(1,3),(2,3),(4,A),(4,D),$ $(5,B),(5,D),(6,C),(6,D),(7,B),(7,C),(8,A),(8,C),$ $(9,A),(9,B)\}$.

The optimal version of the Yu-OH inequality \cite{KBLGC12} can be written as
\begin{widetext}
\begin{equation}
	\text{opt}_3=-\frac{1}{25}\left(\sum_{j\in V_1} \langle A_j \rangle + 2 \sum_{j\in V_2} \langle A_j \rangle + \sum_{(j,k) \in E_1} \langle A_j A_k \rangle + 2 \sum_{(j,k) \in E_2} \langle A_j A_k \rangle - 3 \sum_{(j,k,l) \in T} \langle A_j A_k A_l \rangle \right) \le 1,
\end{equation}
\end{widetext}
where $T=\{(1,4,7),(2,5,8),(3,6,9)\}$.

By choosing observables of the form $A_k=2 \proj{v_k}-\openone$, with
	\begin{align}
		&\ket{v_1}=(1,0,0)^T,\;&\ket{v_8}=(\tfrac{1}{\sqrt{2}},0,\tfrac{1}{\sqrt{2}})^T,\nonumber \\
		&\ket{v_2}=(0,1,0)^T,\;&\ket{v_9}=(\tfrac{1}{\sqrt{2}},\tfrac{1}{\sqrt{2}},0)^T,\nonumber \\
		&\ket{v_3}=(0,0,1)^T,\;&\ket{v_A}=(-\tfrac{1}{\sqrt{3}},\tfrac{1}{\sqrt{3}},\tfrac{1}{\sqrt{3}})^T, \nonumber \\ 
		&\ket{v_4}=(0,\tfrac{1}{\sqrt{2}},-\tfrac{1}{\sqrt{2}})^T,\;&\ket{v_B}=(\tfrac{1}{\sqrt{3}},-\tfrac{1}{\sqrt{3}},\tfrac{1}{\sqrt{3}})^T, \nonumber\\
		&\ket{v_5}=(\tfrac{1}{\sqrt{2}},0,-\tfrac{1}{\sqrt{2}})^T,\;&\ket{v_C}=(\tfrac{1}{\sqrt{3}},\tfrac{1}{\sqrt{3}},-\tfrac{1}{\sqrt{3}})^T, \nonumber\\
		&\ket{v_6}=(\tfrac{1}{\sqrt{2}},-\tfrac{1}{\sqrt{2}},0)^T,\;&\ket{v_D}=(\tfrac{1}{\sqrt{3}},\tfrac{1}{\sqrt{3}},\tfrac{1}{\sqrt{3}})^T, \nonumber\\
		&\ket{v_7}=(0,\tfrac{1}{\sqrt{2}},\tfrac{1}{\sqrt{2}})^T,\;&
	\end{align}
and applying the method described in the main text, we obtain the values the value $25/24 \approx 1.042$ for YO and the value $83/75 \approx 1.107$ for opt$_3$.


\section{Effect of partial distinguishability}


The BS mixes the two spatial modes regardless of the degree of freedom that makes the photons distinguishable. That is,
\begin{equation}
	U_{BS}(\phi,\theta)=\text{exp}\left\{\frac{\theta}{2}\left[e^{-i \phi}(a^\dag b+a_\bot^\dag b_\bot)-e^{i \phi}(b^\dag a+b_\bot^\dag a_\bot)\right]\right\}.
\end{equation} 
For example, suppose that $a = a_H$ annihilates horizontally polarized photons and $a_\bot = a_V$ annihilates vertically polarized photons. Then, we effectively have four different modes, i.e.,
$a^\dag\ket{\rm vac}\equiv \ket{1, 0, 0, 0}$, $a_\bot^\dag\ket{\rm vac}\equiv \ket{0, 1, 0, 0}$, $b^\dag\ket{\rm vac}\equiv \ket{0, 0, 1, 0}$, and $b_\bot^\dag\ket{\rm vac}\equiv \ket{0, 0, 0, 1}$.
 Consider the following input state:
\begin{align}
	\ket{1_H, 1_\eta}=& a^\dag_H \left(\sqrt{1-\eta^2} b^\dag_H+\eta b^\dag_V \right) \ket{\rm vac} \\
	=&\sqrt{1-\eta^2} \ket{1, 1,0, 0}+\eta \ket{1, 0,0,1},
\end{align}
where $\ket{n_{aH},n_{aV},n_{bH},n_{bV}}$ is the state with $n_{aH}$ horizontally polarized photons and $n_{aV}$ vertically polarized photons in the upper mode and $n_{bH}$ horizontally polarized photons and $n_{bV}$ vertically polarized photons in the lower mode. The action of the BS on this state involves $U(\theta, \phi_j)\ket{1_H, 1_H}=\ket{v_j}$, which is the same as Eq.~\eqref{11}, and $U(\theta, \phi_j)\ket{1_H, 1_V}=\ket{\tilde v_j}$ which is
\begin{align}
	\ket{\tilde{v}_j}=&\cos^2\frac{\theta}{2} \ket{1, 0,0,1}-\sin^2\frac{\theta}{2}\ket{0,1, 1,0} \nonumber \\
	&+\frac{\sin\theta}{2} (e^{-i\phi_j}\ket{1,1,0,0}-e^{i\phi_j}\ket{0,0,1,1}).
\end{align}
Therefore,
\begin{equation}
	U(\theta, \phi_j)\ket{1, 1_\eta}=\sqrt{1-\eta^2}\ket{v_j}+\eta\ket{\tilde{v}_j}.
\end{equation}
The joint detection probability is then
\begin{equation}
	p_+=(1-\eta^2)\cos^2\theta+\eta^2\frac{1+\cos^2\theta}{2} 
\end{equation}
and the bunching probability is $p_-=1-p_+$. 



\bibliography{Refs}


\end{document}